# State Transition Block Diagram of the Generalized Maxwell Slip Friction Model

## Kirk Roffi

***Abstract*— Dynamic friction models (DFMs) encode essential information for the simulation and control of systems with friction. Traditionally, DFMs have been published with conceptual block diagrams, promoting clarity and reproducibility in simulation. However, modern DFMs have grown increasingly complex and block diagrams are now rarely presented, limiting accessibility. This letter presents a block diagram representation of the Generalized Maxwell Slip (GMS) friction model, an advanced multi-state DFM capable of simulating a wide range of nonlinear friction phenomena. The diagram can be implemented in the MATLAB-Simulink environment using a Stateflow chart or embedded if-else logic to represent the state transition criteria, but it is not limited to this platform. Closed-loop and open-loop simulations were conducted to verify that the block diagram reproduces non-drifting behavior and stick-slip friction, including benchmarking against the LuGre model. The proposed diagram improves accessibility to advanced dynamic friction models and provides the engineering community with a practical tool for the simulation and control of systems with friction.**

## I. INTRODUCTION

Friction is a major contributor to tracking error in precision motion controls, motivating the pursuit of accurate, model-based friction compensation. Dynamic friction models (DFMs) provide compact mathematical representations of complex friction phenomena, suitable for real-time simulation and control. The enduring legacy of DFMs for controls is exemplified by the Dahl model, which has been applied for over half a century to compensate for hysteretic losses [1], [2].

Dynamic friction models generally adopt empirical, reduced-order frameworks to describe friction phenomena. This is achieved using nonlinear state equations based on an internal variable $z$ which represents the deflection of microscopic surface asperities or bristles [3]. Among DFMs, the LuGre model is perhaps the most widely-used for control applications, employing a single internal state to represent the average bristle deflection [4], [5].

Simulation is a fundamental tool for analyzing DFMs and associated control strategies [6], [7]. To enable simulation, DFMs must be translated from mathematical expressions into executable code or block diagrams. Classical DFMs, such as those of Dahl, Karnopp, and Haessig were published with conceptual block diagrams that map almost directly to simulation environments like MATLAB-Simulink [8], [9], [10]. This tradition promoted accessibility, reproducibility and enabled systematic model comparisons [11], [12], [13].

Modern DFMs have evolved considerably, incorporating multiple internal states and switching logic to more accurately model friction phenomena. Notable examples include the Maxwell Slip model and its variants, extensions of the LuGre and Bristle models, and the Generalized Maxwell Slip (GMS) model [14], [15]. These advanced DFMs are considerably more difficult to implement and simulate. As concluded by Kubas et al., "*It needs to be stressed that the GMS friction model is the latest and most advanced of all the dynamic friction models used in the dynamics of mechanical systems but due to its complexity, it is hardly ever used in calculations.*" [16].

Consequently, the traditional practice of publishing explicit block diagrams of DFMs has nearly disappeared. A telling symptom of this shift is that 'ease of implementation' has become a key criterion by which DFMs are evaluated [17]. As a result, advanced models like the GMS remain largely confined to academic research while practitioners often favor simpler models at the expense of performance.

To address this gap, this letter presents a state transition block diagram for the GMS model. The GMS model is an advanced, multi-state DFM that captures a wide range of friction behavior and is suitable for control applications [18], [19]. MATLAB-Simulink software provides a robust block diagram-based simulation environment for dynamic systems, and is widely-used in academia and industry [20]. Within Simulink, the state transition logic for each GMS friction element is nominally implemented using a Stateflow chart , or alternatively as embedded *if-else* logic. The Stateflow chart provides an intuitive, graphical representation of the state transition logic that is consistent with model-based design practices [21].

Al-Bender et al. developed the GMS model to reproduce fundamental friction behaviors while maintaining consistency with asperity-level physics [19]. The GMS model can be considered a successor to the LuGre model; it captures frictional lag and nonlocal hysteresis without drifting, while offering a more physically tractable structure. It models the

Kirk Roffi is an independent scientist and engineer and IEEE member.
Email: kdr152004@gmail.com

deflection of multiple bristles using parallel Maxwell elements (Fig.1). Unlike the LuGre model which relies on the continuous dynamics of a single $z$ state, the GMS employs multiple friction elements with discrete switching between stick and slip states. This added complexity can pose challenges for online state estimation and numerical simulation, further motivating the development of a block diagram representation [22].

This work continues the tradition of providing block diagram representations of DFMs to enhance accessibility, reproducibility, and learning through simulation [11], [23], [24]. The remainder of the letter is organized as follows. Section II reviews the GMS model and presents the block diagram, including a detailed implementation of the state transition logic. Section III presents simulations for well-known test cases from the literature, including benchmarking against the LuGre model. Section IV concludes the letter.

## II. Generalized Maxwell Slip Friction Model

### A. Overview

The GMS model consists of $i$ Maxwell elements, each characterized by a stiffness $k_i$ and viscoelastic coefficient $\sigma_i$ (Fig.1). The deflection of each element's bristle is described by the state variable $z_i$, and the local friction $F_i$ evolves according to the state dynamics $dz_i/dt$. The state dynamics adopt different forms depending on whether the bristle is sticking or slipping. A Stribeck curve $s(v)$, common to all elements, is incorporated into the slipping dynamics to capture lubrication effects.

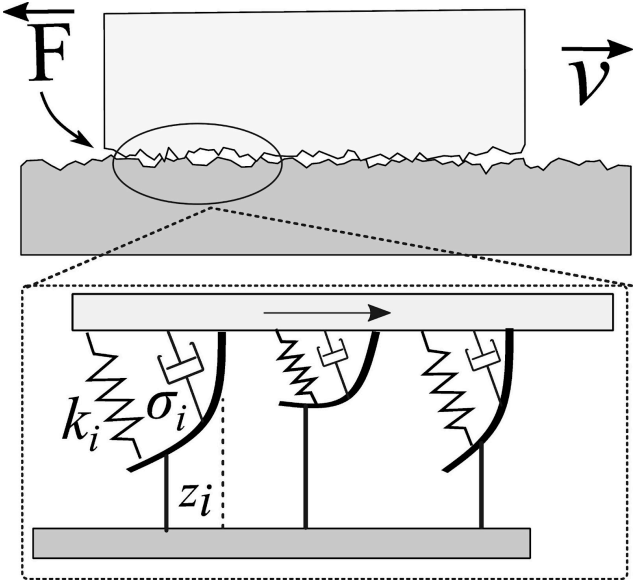

**Fig. 1.** Conceptual illustration of a translational slider with three GMS friction elements.

### B. Mathematical Structure

For each GMS friction element, the state dynamics are described by:

$$\frac{dz_i}{dt} = \left\{ \begin{array}{ll} v & \text{(stick state)} \\ sgn(v)C_i\left(1 - \frac{z_i}{s(v)}\right) & \text{(slip state)} \end{array} \right\} \tag{1}$$

where $v$ is the relative velocity (common to all elements), and $C_i$ is an attraction parameter that governs how fast $z_i$ converges to $s(v)$. The transition between stick and slip states is determined by the comparison of $z_i$ to $s(v)$, and $v$ to zero or a critical velocity $v_c$. The system typically starts in the stick state and transitions to slip once $z_i \geq s(v)$, and transitions back to sticking once $v \leq v_c$. The critical velocity $v_c$ is introduced in this work to relax the numerical stiffness of the state switching, as compared to the original formulation based on strict zero-crossing [19]. A typical Stribeck curve $s(v)$ is given by:

$$s(v) = F_c + \left(F_s - F_c\right)e^{-\left|v/v_s\right|^2} \tag{2}$$

where $F_C$ is the Coulomb friction, $F_s$ is the stiction force, and $v_s$ is the Stribeck velocity. Each element's local friction force $F_i$ is given by:

$$F_i = k_i z_i + \sigma_i \frac{dz_i}{dt} \tag{3}$$

where $k_i z_i$ represents the elastic force and $\sigma_i dz_i/dt$ represents the microviscous force. Then, the total friction force $F$ is given by:

$$F = \sum_{i=1}^{n} F_i + \sigma_2 v \tag{4}$$

where $\sigma_2$ represents the viscous friction coefficient which is common to all elements.

### C. Block Diagram

The block diagram representation is provided in Fig.2a. The slider velocity $v$ is used to compute $s(v)$ using a MATLAB function block in Simulink which encodes equation (2). Then, $v$, $s(v)$, and $z_i$ are input to the Stateflow chart which contains the state transition logic (Fig.2b). Within the Stateflow chart, $dz/dt$ and $F_i$ are computed per equations (1) and (3) respectively, depending on whether each element is in the stick to slip state. At each time step of the simulation, $dz/dt$ is integrated in Simulink (outside the Stateflow chart) to update $z_i$. During the simulation, $dz/dt$ and $z_i$ are updated iteratively for each GMS friction element, as indicated by the feedback loops in Fig.2a. The total friction force $F$ is obtained by summing all local forces $F_i$ and the viscous friction term per equation (4).

### D. Stateflow Chart

The state transition logic for each GMS friction element is nominally represented using a Stateflow chart as a subsystem within the block diagram (Fig.2b). Since the GMS switching rules are relatively simple, the chart is functionally comparable to the equivalent if-else implementation (Fig.2c) as discussed in Section III. However, Stateflow is generally preferred for friction modeling because it provides a clear and concise representation of the state transition logic [24], [25], [26].

Within the Stateflow chart (Fig.2b), the vertical arrow initializes each element in the sticking state, representing a bristle at rest, and can be customized for each element. In the lower-level code (Fig.2c), the initial bristle state is specified as a local initial condition. For both implementations, the stick to slip logic compares $z$ and $s(v)$ in both positive and negative directions to ensure alignment between the direction of bristle micro-motion and the slider macro-motion. Conversely, the slip to stick logic is based on the absolute values of $v$ and $s(v)$, allowing bristles to stick regardless of the sliding direction. The inclusion of both $z$ vs. $s(v)$ and $v$ vs. $v_c$ in the transition criteria helps prevent non-physical edge cases, for example, if a bristle snapping because $z$ exceeds $s(v)$ but the slider remains below the critical velocity. For edge cases where $z_i=s(v)$, the system transitions from stick to slip, which is consistent with the original formulation of the GMS model [19].

While the Stateflow implementation is not fully vectorized for large numbers of elements, this limitation is acceptable since only four to twelve GMS elements are usually sufficient for fitting experimental data [19], [27], [28]. An advantage of the Stateflow chart, compared to lower-level code, is the ability to visualize each element's active state in real-time during simulation. This capability is especially useful for debugging advanced stick-slip models with more than two states, such as those presented by Nevarro-Lopez [26].

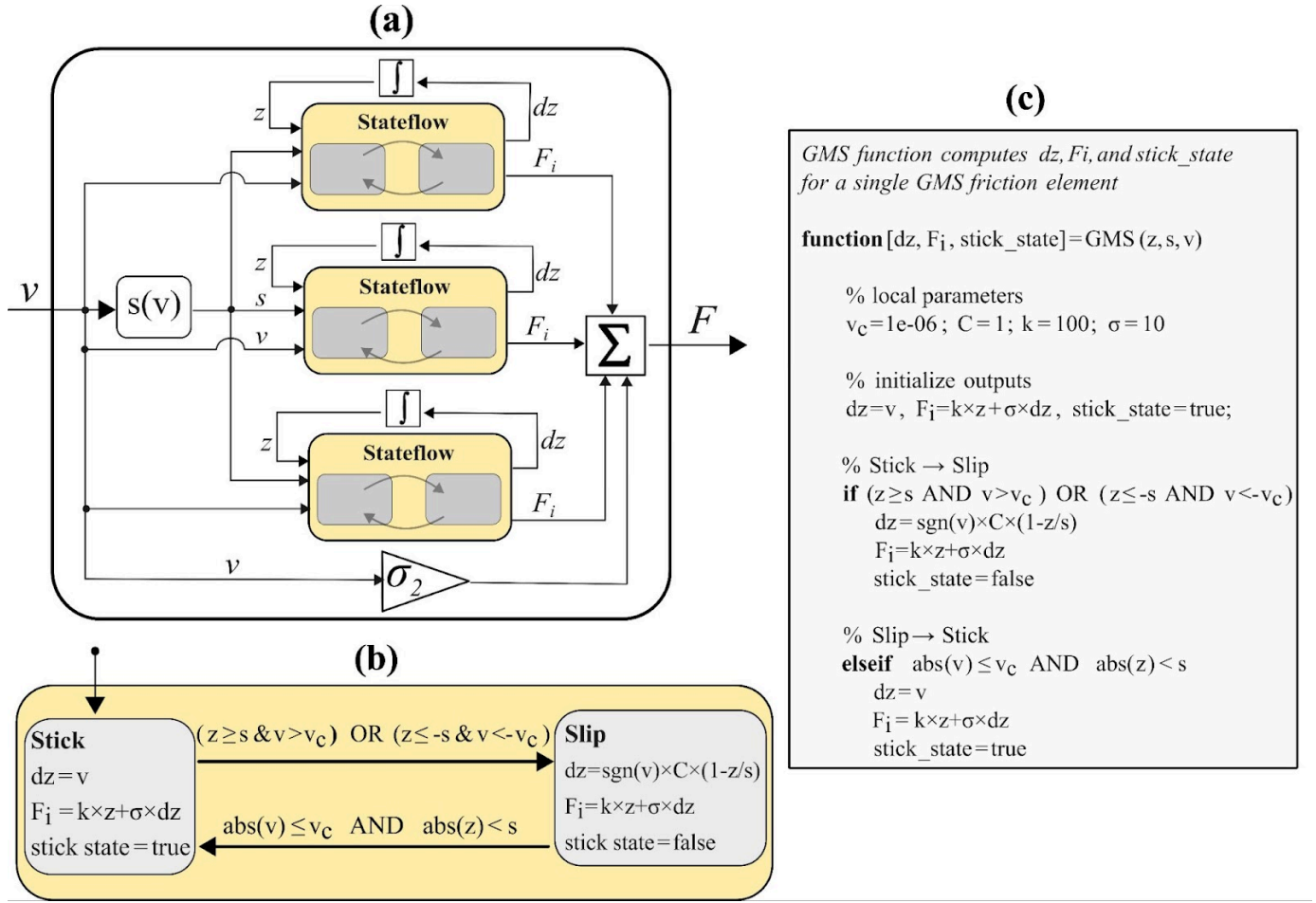

**Fig. 2.** a) Block diagram representation of the GMS model with *i=3* friction elements, b) Stateflow chart for the state transition logic of a single friction element, c) equivalent if-else pseudo-code for comparison.

## III. SIMULATION OF GMS FRICTION

This section applies the block diagram to simulate GMS friction in both closed-loop and open-loop configurations for well-known test cases from the literature, including non-drifting and stick-slip behaviors. For benchmarking, simulations using the Stateflow chart (Fig.2b) are compared with the equivalent if-else logic encoded using a MATLAB function block in Simulink (Fig.2c). As an additional benchmark, the GMS model is compared with the LuGre model.

All simulations were performed in MATLAB-Simulink R2025b using the ode45 solver with a maximum step size of $10^{-3}$ s and a relative tolerance of $10^{-4}$. The solver setting for shape preservation was set to 'disable all' and the zero-crossing algorithm was set to 'adaptive'. The GMS model parameters were manually tuned to elicit comparable behavior as the LuGre model, and are summarized in Table I for a slider with a mass of 1 kg. All $z$ state integrators were initialized to zero. The LuGre parameters were adapted from Canudas de Wit et al., using an average bristle stiffness of $10^3$ N/m and a viscoelastic coefficient of 31.6 Ns/m [4].

### A. Non-Drifting Simulation

A distinguishing feature of the GMS model is its non-drifting property [3], [19]. Drifting occurs in response to noisy inputs and is an unintended artifact for certain DFMs that capture pre-sliding hysteresis without accurately rendering stiction [29]. To test this behavior, an applied force was designed to ramp-up to a value less than the break-away and then oscillate sinusoidally to mimic sensor noise [29]. In this closed-loop simulation, friction is fed back to influence the slider dynamics. The results are summarized in Fig.3.

As the applied force increased, the GMS and LuGre models both exhibited linearly elastic friction. When the applied force decreased, both models showed hysteresis and the friction force traversed towards the origin along a new trajectory. However, at the onset of sensor noise, the models diverged; the LuGre model showed steady positional drift, confirming the drifting phenomenon (Fig.3d). In contrast, the GMS model produced bounded oscillatory micro-motion within the initial hysteresis loop (Fig.3b and 3c). These results confirm that the block diagram faithfully reproduced the non-drifting behavior of the GMS model.

Non-drifting behavior was generally consistent across different implementations of the state transition logic (Fig.3b and 3c). This result is expected since the noisy input did not trigger state transitions in any of the GMS friction elements. However, when the applied force oscillated, small differences in friction ($\sim 10^{-4}$ N) and position ($\sim 10^{-6}$ m) were detected between implementations. These discrepancies may stem from subtle interactions between the logic-containing subsystems and the Simulink solver, such as execution priority differences between Stateflow charts and MATLAB function blocks. Such sub-micron variations of $10^{-6}$ m are likely tolerable since feedforward tracking error for the GMS model has been reported on the order of $10^{-3}$ m [27].

### GMS MODEL PARAMETERS

| Parameter | Non-Drifting Simulation | Stick-Slip Simulation | Unit |
|---|---|---|---|
| $i$ | 4 | 4 | - |
| $k_i$ | [100, 10, 1, 0.1] | [0.7, 0.7, 0.7, 0.7] | N/m |
| $\sigma_i$ | [10, 0.1, 1, 1] | [0.3, 0.4, 0.5, 0.6] | Ns/m |
| $C_i$ | [1, 1, 1, 1] | [0.1, 0.05, 0.025, $10^{-4}$] | m/s |
| $\sigma_2$ | 4 | 0.1 | Ns/m |
| $F_S$ | 1.5 | 1.5 | N |
| $F_C$ | 1 | 0.4 | N |
| $v_s$ | 0.001 | 0.001 | m/s |
| $v_c$ | $10^{-6}$ | $10^{-6}$ | m/s |

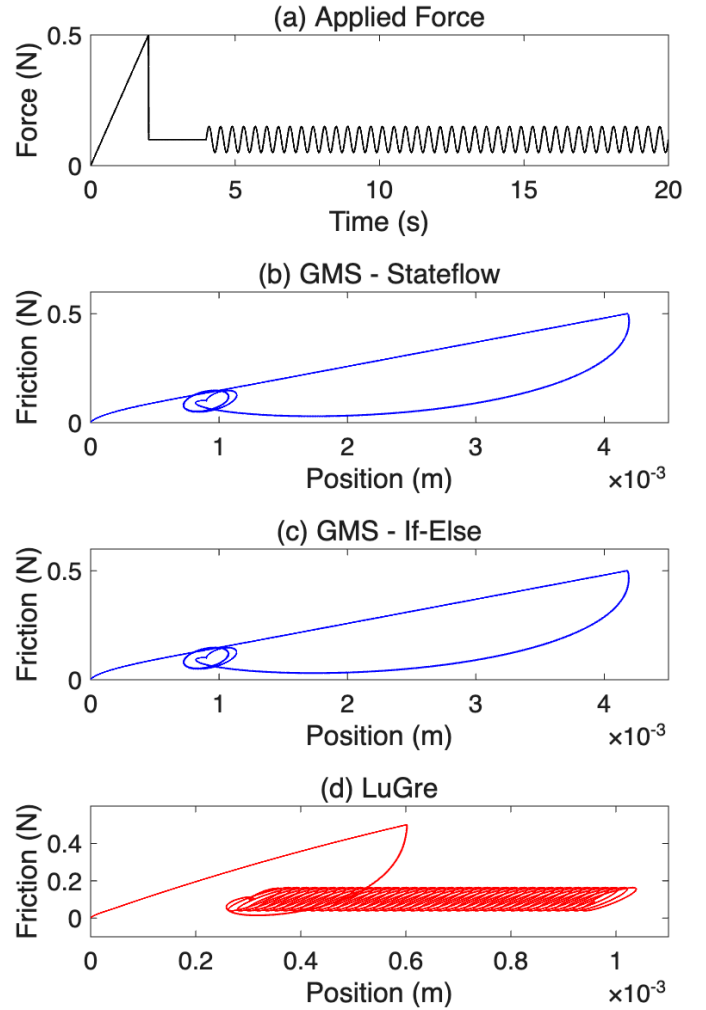

**Fig. 3.** Non-drifting simulation results: a) applied force, b) GMS friction (Stateflow implementation), c) GMS friction (if-else implementation), d) LuGre friction

## B. Stick-Slip Simulation

The GMS and LuGre models differ fundamentally in how they represent stick-slip friction i.e. transitions between static and dynamic friction [19]. In the GMS model, stick-slip is modelled explicitly using a discrete switching mechanism, while in the LuGre model, stick-slip emerges implicitly from the continuous dynamics of the $z$ state. Unlike the LuGre model which uses one internal state to represent the average bristle deflection, the GMS model assigns a unique $z$ state to each friction element. By distributing friction across multiple internal states, the GMS model can capture spatial heterogeneity among surface asperities, potentially enhancing physical realism.

To simulate stick-slip, a sinusoidal velocity signal was designed based on the approach of Jankowski et al. and input directly to the GMS and LuGre models (Fig.4a) [30]. Simulations were performed in an open-loop configuration whereby friction was not fed back into the slider dynamics. The results are summarized in Fig.4.

During the first stick-slip cycle, the GMS internal states were aligned (Fig.4b) in a similar manner as the LuGre's single internal state (Fig.4c), resulting in a singular break-away peak near $t=5s$ (Fig.4d and 4e). Beyond the first cycle, the GMS model produced complex stick-slip effects, owing to the diversity of bristle parameters in Table I, and driven by divergence among the $z$ states (Fig.4b).

The noncohesive $z$ states led to two key behaviors not captured by the LuGre model: 1) asymmetric friction depending on the velocity direction, and 2) multiple break-away peaks (Fig.4d and 4e). The asymmetry arises because integration of the $z$ states does not reset when state transitions occur i.e. the integrator does not fully unwind between changes in sliding direction. The unique capability of the GMS model to capture multiple break-away peaks is consistent with experimental observations, whereby micro-slip events can precede or coincide with the transition to macroscopic slipping (Fig.4d and 4e) [31].

In contrast to the GMS model, the LuGre model produced more uniform stick-slip behavior i.e. it did not exhibit multiple break-aways or asymmetry (Fig.4f). This result is expected since the single $z$ state is based on the average bristle deflection, and the LuGre is not parameterized to model heterogeneous bristles [4]. Moreover, since the state dynamics do not involve discrete switching, the state integrator fully unwinds between cycles, preventing directional bias.

As in the non-drifting simulations, stick-slip behavior was qualitatively consistent across different implementations of the state transition logic (Fig.4d and 4e). However, close inspection of the residual, i.e. the difference in friction between the Stateflow and if-else implementations, revealed discrepancies on the order of $10^{-3}$ N, with occasional spikes coinciding with state transitions. This indicates that the two implementations are qualitatively similar but not quantitatively identical, suggesting that model validation should carefully consider which implementation most accurately reflects experimental observations.

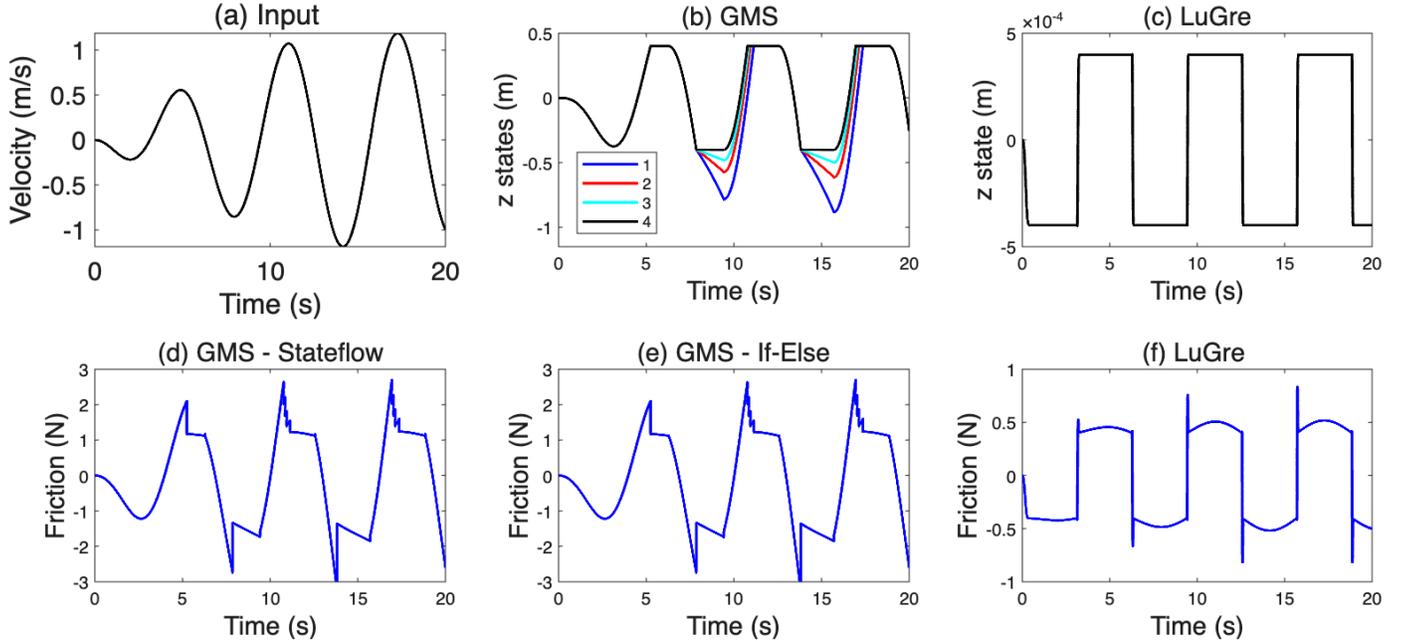

**Fig. 4.** Stick-slip simulation results: a) velocity input, b) GMS internal $z$ states ($i=4$) , c) LuGre $z$ state, d) GMS friction (Stateflow implementation) e) GMS friction (if-else implementation), f) LuGre friction

## IV. Conclusion

This letter presents a state transition block diagram representation of the Generalized Maxwell Slip friction model. Simulation results demonstrate that the block diagram faithfully reproduces key model behaviors, including non-drifting and asperity-level stick-slip effects, with benchmarking against the LuGre model. The proposed diagram improves accessibility to advanced dynamic friction models and provides the engineering community with a practical tool for the simulation and control of systems with friction.

## Acknowledgment

This work was conducted independently as a self-funded project in a personal, non-commercial context. The views expressed herein are my own and do not reflect those of my employer.